\documentclass[twocolumn,showpacs,showkeywords,amsmath,amssymb]{revtex4}
\usepackage{graphicx}
\begin{document}
\title{Equation of State for Neutralino Star as a Form of Cold Dark Matter}
\author{Jie Ren}
\email{jrenphysics@hotmail.com}
\author{Xue-Qian Li}
\email{lixq@nankai.edu.cn}
\author{Hong Shen}
\email{songtc@nankai.edu.cn} \affiliation{Department of physics,
Nankai University, Tianjin 300071, China}
\date{\today}
\begin{abstract}
In order to study the structure of neutralino star and dark galaxy,
we consider dynamical interactions due to boson-exchange in the
neutralino matter. Taking into account interactions of neutralinos
with bosons, we derive the equation of state (EOS) of neutralino
stars in terms of the relativistic mean field approach. Then we
apply the resulting EOS to investigate properties of the neutralino
star such as its density profile and mass limit. For example, if the
neutralino mass is around 1 TeV, the Oppenheimer mass limit of the
neutralino star is obtained as $6.06\times 10^{-7}M_\odot$, and the
corresponding radius is about $7.8$ mm. Actually, due to an
increasing annihilation rate as indicated by our calculation, this
dense state can never be realized in practice. Our results also show
that the low density neutralino star may be a possible aggregation
of the cold dark matter.
\end{abstract}
\maketitle

\section{Introduction}
The SNe Ia and WMAP data \cite{rie04,spe06} in observational
cosmology indicate that our Universe is composed of approximately
$4\%$ ordinary matter, $22\%$ dark matter and $74\%$ dark energy. To
explain the nature of the cold dark matter is one of the most
challenging tasks in modern physics. Meanwhile, theorists take great
efforts to find new physics beyond the standard model (SM). The
supersymmetry \cite{hab85} is believed to be the most possible
extension of the SM. Within the framework of the minimal
supersymmetric standard model (MSSM), the lightest neutralino
$\tilde{\chi}_1^0$ is a natural and attractive candidate for the
cold dark matter \cite{jun95}. Annihilation of the neutralinos
provides some detectable signals, such as the high energy gamma rays
\cite{for04,ber00}. The authors of Refs.~\cite{gur95,gur96,gur97}
proposed a model of ``neutralino star" as a gravitationally
compressed object of non-baryonic cold dark matter. They calculated
the diffuse galactic and extragalactic gamma radiation and claimed
that the results are in reasonable agreement with observations. They
used some phenomenological formulae to describe the density
distribution in the neutralino star. Following the literature, we
are going to study the properties of the neutralino star by
concerning the dynamical interactions among neutralinos. The concept
of ``neutralino star" in this paper represents a compact star that
only consists of neutralinos, and it is slightly different from the
picture of Refs. \cite{gur95,gur96,gur97}.

The hierarchical structure of dark matter should be investigated if
we want to know the nature of the dark matter. Physical mechanisms
have been proposed for the formation of dark matter objects in
literature. Gurevich \textit{et al.} studied how far does the
hierarchical structure of dark matter go for small scales
\cite{gur95}. The scale length of the created object is determined
by the moment when the Jeans instability reaches the nonlinear stage
\cite{gur95}. They also show that the dark matter particles can be
self-trapped by gravitational forces and form dark matter stars
\cite{gur88}. Kolb and Tkachev studied the nonlinear dynamics for
axions to form boson stars \cite{kol93,kol94}. Matos \textit{et al.}
studied the scalar dark matter \cite{mat00,alc03} and examined the
possibility that galactic halos are collisionless ensembles of
scalar ``massive compact halo objects" whose mass is in the range
$m<10^{-7}M_\odot$ or $30M_\odot<m<100M_\odot$ \cite{her04}. Later
we will show that the mass of the neutralino star also falls in this
range. Neutralinos as stable dark matter can also aggregate to form
structures, thus, the self-annihilation and aggregation processes
caused by gravity may give rise to a dynamical equilibrium, and the
resultant equilibrium would lead to a formation of the dark matter
star.

Relativistic mean field (RMF) approach \cite{ser86} has been widely
used in nuclear physics and particle physics with dense medium, and
it provides a successful description of nuclear matter and finite
nuclei, including unstable nuclei. It has also been applied to
construct the equation of state (EOS) for neutron stars and
supernovae. Therefore one can believe that it is applicable to our
case. Starting with an effective Lagrangian of nucleon-meson
interactions, one can determine the EOS of the neutron star by using
the RMF theory, and then apply it to study neutron star properties
\cite{she98a,she98b,she02}. The situation of the neutralino star is
similar, as both nucleons and neutralinos are stable fermions except
neutrons do not annihilate each other, but neutralonos do. Thus, it
is quite probable that the neutralino star stays in an equilibrium
of the forces caused by the gravitation and the degenerate pressure
and mutual interactions. In this scenario neutralinos can interact
with each other by exchanging neutral Higgs bosons and $Z$ boson,
just as the mutual interaction between nucleons is mediated by
mesons. Narain \textit{et al.} studied the compact stars made of
fermionic dark matter with interactions, which contribute an
additional term in the EOS determined with the ideal Fermi gas
model, as a first order approximation \cite{nar06}.

In this work, with the neutralino-boson interaction, we use the RMF
approach to construct the EOS of the neutralino star, and then
investigate its properties. By solving the Oppenheimer-Volkoff (OV)
equation, the mass density distribution is obtained. For simplicity,
we assume that the constituents of neutralino star are only the
lightest neutralinos, which are interacting with each other by
exchanging four MSSM bosons, and adopt a parameter set for the
numerical computation, which is carried out in terms of the program
DarkSUSY 4.1 \cite{gon04}. The mass of the neutralino star as a
function of its central mass density is given and this result is
consistent with the OV limit calculated with the ideal Fermi gas
model \cite{bil98}. The result of the dynamical simulations of the
neutralino star shows that the neutralino star is a possible
existent form of the cold dark matter, especially in the dark
galaxy.

The paper is organized as follows: In the next section, we use the
RMF approach to construct the EOS of the neutralino star. In Sec.
III, we apply the EOS to study the properties of neutralino star.
The last section is devoted to our conclusion and the discussion on
dark galaxy.

\section{Relativistic mean field theory for neutralino matter}
Within the framework of MSSM, the neutralino can interact with four
bosons, $H^0$, $h^0$, $A^0$, and $Z$. Here $H^0$ and $h^0$ are the
two CP-even neutral Higgs bosons, $A^0$ is the CP-odd neutral Higgs
boson, and $Z$ is a vector boson in the SM. The effective Lagrangian
can be written as
\begin{eqnarray}
\mathcal{L} &=& \bar{\chi}(i\gamma^\mu\partial_\mu-m_\chi-g_H H-g_h
h-g_A\gamma_5 A-g_Z\gamma^\mu Z_\mu)\chi\nonumber\\
&& +\frac{1}{2}\partial_\mu H\partial^\mu H-\frac{1}{2}m_H^2
H^2+\frac{1}{2}\partial_\mu h\partial^\mu h-\frac{1}{2}m_h^2 h^2\nonumber\\
&& +\frac{1}{2}\partial_\mu A\partial^\mu A-\frac{1}{2}m_A^2 A^2
-\frac{1}{4}F_{\mu\nu}F^{\mu\nu}+\frac{1}{2}m_Z^2 Z_\mu Z^\mu,
\end{eqnarray}
where $\chi$, $H$, $h$, $A$, and $Z^\mu$ denote the fields of
$\tilde{\chi}_1^0$, $H^0$, $h^0$, $A^0$, and $Z$, respectively, and
$F_{\mu\nu}\equiv \partial_\mu Z_\nu-\partial_\nu Z_\mu$.

Starting with this Lagrangian, a set of Euler-Lagrange equations can
be derived. The Dirac-type equation for the neutralino field is
given as
\begin{equation}
(i\gamma^\mu\partial_\mu-m^*-g_Z\gamma^\mu Z_\mu)\chi=0,
\end{equation}
where $m^*=m_\chi+g_H H+g_h h+g_A\gamma_5A$ is the effective mass of
neutralino. The field equations for $H^0$, $h^0$, $A^0$, and $Z$ are
given as
\begin{eqnarray}
\partial_\mu\partial^\mu H+m_H^2 H &=& -g_H\bar{\chi}\chi,\\
\partial_\mu\partial^\mu h+m_h^2 h &=& -g_h\bar{\chi}\chi,\\
\partial_\mu\partial^\mu A+m_A^2 A &=& -g_A\bar{\chi}\gamma_5\chi,\\
\partial_\nu F^{\mu\nu}+m_Z^2 Z^\mu &=& g_Z\bar{\chi}\gamma^\mu\chi.
\end{eqnarray}
These equations are coupled partial differential equations, which
are difficult to be solved exactly. To get an approximate solution
which can properly describe the physical picture, we adopt the
relativistic mean field approximation \cite{ser86}. In this approach
the boson fields are treated as classical fields, and the field
operators $H$, $h$, $A$, and $Z^\mu$ are replaced by their
expectation values. Here we consider an approximation that the
bosons are treated as static infinite matter to obtain simplified
equations, where derivative terms in the Klein-Gordon equations
vanish automatically due to the uniformity of the matter. The
spatial components of the vector field $Z^\mu$ vanish since there is
no vector current in average, i.e.,
$\langle\bar\chi\gamma^i\chi\rangle=0$. Hence, the equations for the
boson fields are reduced to
\begin{eqnarray}
H &\equiv& \langle H\rangle=-\frac{g_H}{m_H^2}\langle\bar{\chi}\chi\rangle,\\
h &\equiv& \langle h\rangle=-\frac{g_h}{m_h^2}\langle\bar{\chi}\chi\rangle,\\
A &\equiv& \langle
A\rangle=-\frac{g_A}{m_A^2}\langle\bar{\chi}\gamma_5\chi
\rangle,\\
Z &\equiv& \langle
Z^0\rangle=\frac{g_Z}{m_Z^2}\langle\bar{\chi}\gamma^0 \chi\rangle.
\end{eqnarray}
At zero temperature, the equations of the boson fields can be
written as
\begin{eqnarray}
H &=& -\frac{g_H}{m_H^2\pi^2} \int_0^{k_F} \frac{m^*}{\sqrt{k^2+{m^*}^2}}k^2
d k,\\
h &=& -\frac{g_h}{m_h^2\pi^2} \int_0^{k_F} \frac{m^*}{\sqrt{k^2+{m^*}^2}}k^2
d k,\\
A &=& 0,\\
Z &=& \frac{g_Z}{3m_Z^2\pi^2}k_F^3,
\end{eqnarray}
where $k_F$ is the Fermi momentum of neutralino. Indeed, the
temperature in neutralino star is much lower than any concerned
energy scale, thus it is safe to take the zero-temperature
approximation. The pseudo-scalar Higgs $A^0$ has no contribution in
the RMF approach. These equations should be solved
self-consistently, then the energy density and the pressure can be
obtained from the energy momentum tensor~\cite{ser86}. The energy
density of neutralino matter is given by
\begin{equation}
\rho=\frac{1}{\pi^2}\int_0^{k_F}\sqrt{k^2+m^{*2}}k^2d k
+\frac{1}{2}m_H^2 H^2+\frac{1}{2}m_h^2 h^2+\frac{1}{2}m_Z^2 Z^2,
\end{equation}
and the pressure is given by
\begin{equation}
p=\frac{1}{3\pi^2}\int_0^{k_F}\frac{k^4d k}{\sqrt{k^2+m^{*2}}}
-\frac{1}{2}m_H^2 H^2-\frac{1}{2}m_h^2 h^2+\frac{1}{2}m_Z^2
Z^2.\label{eq:p}
\end{equation}
To obtain a relation between $\rho$ and $p$ from $\rho=\rho(k_F)$
and $p=p(k_F)$, we numerically vary the value of $k_F$ in a wide
range, then construct the EOS of the neutralino matter.

\begin{figure}[]
\centering
\includegraphics{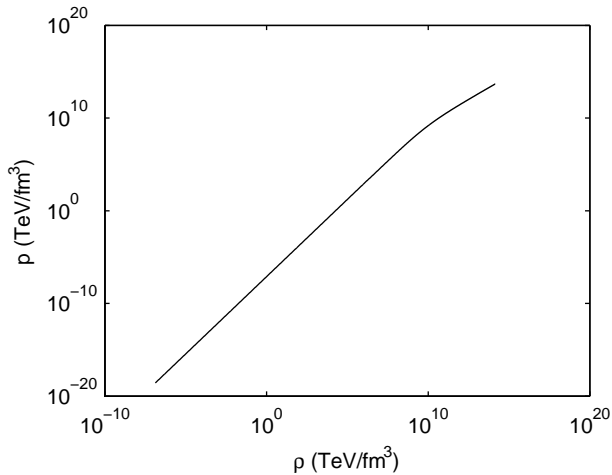}
\caption{\label{fig1} The EOS of the neutralino matter.
$\rho$ and $p$ denote the energy density and the pressure, respectively.}
\end{figure}

There are several available models in the program DarkSUSY
\cite{gon04} with simplified MSSM parameterization for studies on
supersymmetric dark matter, and here we use the model JE56A\_003796.
Table~\ref{tab:t1} presents our input parameters for the MSSM and
Table~\ref{tab:t2} shows the parameters in the Lagrangian. The EOS
of the neutralino matter in the energy density range $10^{-6}\sim
10^{14}$ TeV/fm$^3$ is shown in Fig.~\ref{fig1}.
\begin{table}[h]
\caption{\label{tab:t1} Input parameters for the MSSM sector}
\begin{ruledtabular}
\begin{tabular}{cc}
Parameter & Value\\
\hline
$\mu$ & $1020.2370$\\
$M_2$ & $2894.9567$\\
$m_A$ & $753.24107$\\
$\tan\beta$ & $14.672504$\\
$m_s$ & $4001.3887$\\
$A_t/m_s$ & $-0.57518824$\\
$A_b/m_s$ & $0.62302097$\\
\end{tabular}
\end{ruledtabular}
\end{table}
\begin{table}[h]
\caption{\label{tab:t2} The parameter set for the Lagrangian}
\begin{ruledtabular}
\begin{tabular}{cc}
Parameter & Value\\
\hline
$m_\chi$ (GeV) & $1016.79763$\\
$m_H$ (GeV) & $753.36438$\\
$m_h$ (GeV) & $118.760345$\\
$m_A$ (GeV & $753.24107$\\
$m_Z$ (GeV) & $91.187$\\
$g_H$ & $0.3131$\\
$g_h$ & $-0.02705$\\
$g_A$ & $-0.02726i$\\
$g_Z$ & $0.00143$\\
\end{tabular}
\end{ruledtabular}
\end{table}

\section{Properties of the neutralino star}
Applying the EOS of the neutralino matter, we calculate the
neutralino star density profile by solving the OV equation
\begin{equation}
\frac{dp}{dr}
=-\frac{GM\rho}{r^2}\left(1+\frac{p}{\rho}\right)\left(1+\frac{4\pi
r^3p}{M}\right)\left(1-\frac{2GM}{r}\right)^{-1},\label{eq:ov}
\end{equation}
where
\begin{equation}
M(r)=\int_0^r4\pi r'^2\rho(r')dr'.
\end{equation}
For a given central mass density, which is related to the Fermi
momentum $k_F$, we solve Eq.~(\ref{eq:ov}) from inside to outside
until the pressure vanishes, then the density profile and the mass
of the neutralino star can be obtained.

\begin{figure}[]
\centering
\includegraphics{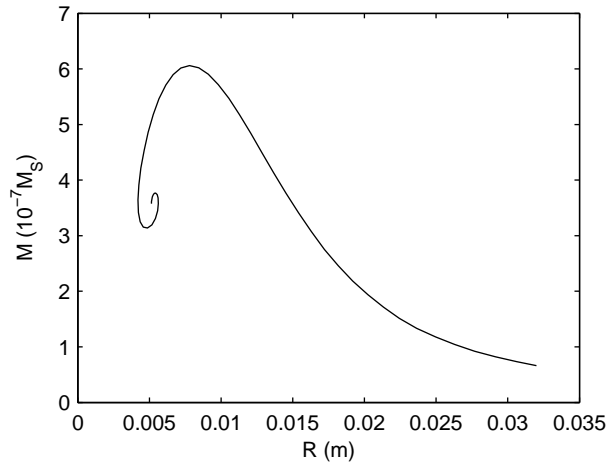}
\caption{\label{fig2} The mass-radius relation of the neutralino
star. $M_{\rm S}$ denotes the mass of the sun.}
\end{figure}

\begin{figure}[]
\centering
\includegraphics{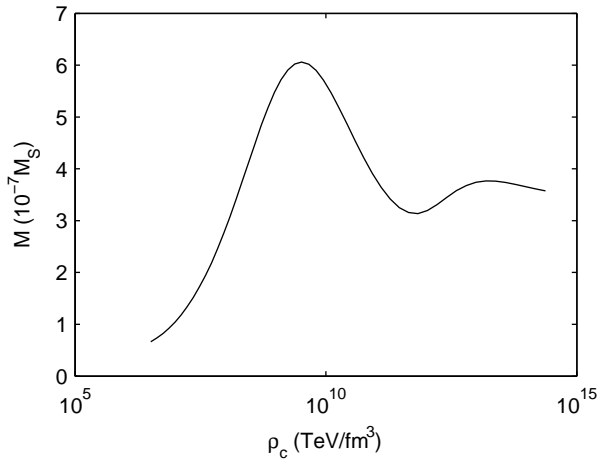}
\caption{\label{fig3} The mass of the neutralino star as a function
of its central mass density. $M_{\rm S}$ denotes the mass of the
sun.}
\end{figure}

\begin{figure}[]
\centering
\includegraphics{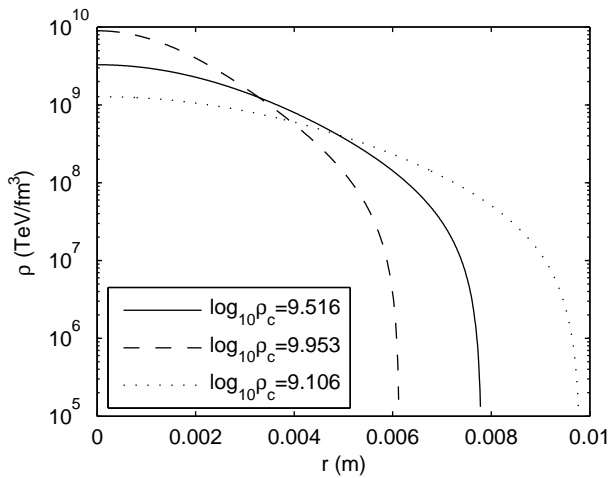}
\caption{\label{fig4} The density profile of the neutralino star.
The three lines corresponds to three different central mass
densities.}
\end{figure}

We take the neutralino mass around 1 TeV for example to perform the
calculations. The mass of the neutralino star as a function of its
radius is illustrated in Fig.~\ref{fig2}, which shows that the mass
limit of the neutralino star is $6.06\times 10^{-7}M_\odot$ (It is
less than the mass of the earth.). Also, the radius of a neutralino
star whose mass just takes the mass limit is only $7.8$ mm. The mass
of the neutralino star as a function of its central mass density is
illustrated in Fig.~\ref{fig3}, and the density profile is plotted
in Fig.~\ref{fig4}.
\begin{figure}[]
\centering
\includegraphics{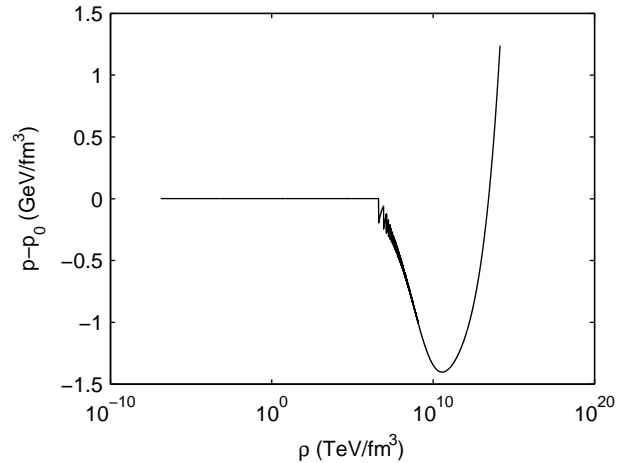}
\caption{\label{fig5} The difference between the EOS including
interactions ($p$) and the one without interactions ($p_0$) for
neutralino matter.}
\end{figure}

By assuming that a compact object of constituents is described by
the ideal Fermi gas model, the OV limit is calculated in
Ref.~\cite{bil98},
\begin{equation}
M_{\rm OV}=2.9924\times 10^9 M_\odot\left(\frac{17.2
\textrm{keV}}{mc^2}\right)^2 g^{-1/2},
\end{equation}
where $m$ is the mass of the possible constituents (e.g., neutron or
neutralino) which form the compact star, and $g$ is the degeneracy
factor, i.e., $g=2$ for Majorana particles and $g=4$ for Dirac
particles. By this formula, the mass limit of the neutralino star
without interactions is $M_{\rm OV}=6.05\times 10^{-7}M_\odot$, and
the limit calculated in our approach is consistent with this OV
limit. If we set the couplings to be zero, then the EOS reduces to
that for ideal Fermi gas of neutralinos. Fig.~\ref{fig5} plots the
difference for the EOS with and without the interactions among
neutralinos, and it shows that the effect of the concerned
interactions is rather small, namely, the gravity and the degenerate
pressure are overwhelmingly important to responsible for the
formation of neutralino stars. From Eq.~(\ref{eq:p}), we can see
that the scalar Higgs bosons $H^0$ and $h^0$ have negative
contributions to the pressure and provide an attractive potential
($V_{\rm A}$), while the vector gauge boson $Z$ has a positive
contribution to the pressure and provides a repulsive potential
($V_{\rm R}$). Obviously, in the low density region, the
interactions can be neglected. As the density goes larger, it
becomes $V_{\rm A}>V_{\rm R}$, and when the density is large enough,
$V_{\rm R}>V_{\rm A}$ finally.

\section{Conclusion and discussion}
In summary, we have obtained the EOS of neutralino matter at zero
temperature in the density range from $10^{-6}$ to $10^{14}$
TeV/fm$^3$ and employed the present EOS to calculate the neutralino
star properties for a specific choice of MSSM parameters. We assume
that the compact neutralino star only consists of the neutralino
matter and we use the RMF approach, which is obvious a reasonable
approximation in our case. For example, if the neutralino mass is
around 1 TeV, the Oppenheimer mass limit of the neutralino star is
$6.06\times 10^{-7}M_\odot$ and the corresponding radius is $7.8$
mm. From the other side, there exist two opposite processes, first,
the gravity causes the neutralinos at vicinity close to a core to
collapse into a small region as the more matter aggregates, the
denser the star becomes, but meanwhile as the star becomes denser,
the annihilation rates of neutralinos would rapidly increase and
neutralinos annihilate into SM particles, which would escape from
the star. The process prevents more neutralinos to collapse into a
core. The final state of the neutralino star would be a result of a
competition of the two processes and in our later work we are going
to evaluate the resultant densities which would determine the size
and average mass of such neutralino stars. The resultant stars might
be a candidate for main components of a dark galaxy.

One is tempted to postulate that the neutralino star is a possible
existent form of the cold dark matter, especially the relatively low
density neutralino star in dark galaxy. VIRGOHI 21, a dark galaxy in
the Virgo cluster, has been discovered recently \cite{min05}. With
belief on validity of supersymmetry, it is reasonable to assume that
the neutralino matter might be the main content of the dark galaxy.
The high energy gamma ray as detectable signals produced by dark
matter is intensely studied in the literature \cite{bi05,ber03}. The
central density of a neutralino star near its mass limit is around
$10^{52}$ GeV/cm$^3$, $53$ orders of magnitude higher than the local
dark matter density of smooth profile. If the compact neutralino
star were a main component in the dark galaxy, not only its
annihilation would dramatically change the profile of the dark
galaxy, but also the gamma ray flux would contradict the
experimental data. However, we should not exclude the possibility of
the existence of a few neutralino stars especially low density ones.
For example, the mean density of the dark galaxy can be 0.3
GeV/cm$^3$ by choosing an appropriate central density of such stars,
which is reasonable to explain the observed gamma ray flux. We think
that the dark matter can accrete on the neutralino star due to its
strong gravitational force, thus the dynamical equilibrium of its
self-annihilation and the dark matter accretion will cause such
stars to live longer. Therefore, the compact neutralino star near
its mass limit should not exist in the dark galaxy, whereas a few
small mass neutralino stars are a possible aggregation form of dark
matter.

As a further study on the interactions among the neutralino matter
more precisely, we should consider not only the lightest neutralino
$\tilde{\chi}_1^0$, but also the heavier ones
$\tilde{\chi}_{2,3,4}^0$, as well as the charginos
$\tilde{\chi}_{1,2}^\pm$ and other possible particles mixed in the
neutralino star, even though they may decay into lightest neutralino
and SM particles. The axion is also an interesting candidate to
study \cite{kol93,kol94}. Generally, for a mixture of neutralinos
with bosons or heavier fermions the mass limit would be smaller and
a mixing with lighter fermions will make the mass limit larger. In
addition, the similar method can also be applied to the case of the
``neutrino star" \cite{bil98}, thus, the properties of the neutron
star, neutralino star and neutrino star all can be studied in a
unique theoretical framework.

\section*{Acknowledgements}
We thank Prof. X.H. Meng and Prof. X.J. Bi for helpful discussions,
and Prof. P. Gondolo for helps of the program DarkSUSY. This work is
partly supported by the National Natural Science Foundation of China
(NNSFC).

\end{document}